\documentclass[11pt,onecolumn, draftcls]{IEEEtran}

\usepackage{epsfig,amsfonts,amsmath,graphicx,epsfig}
\usepackage{subfigure,graphics,amssymb,amsxtra,color}
\usepackage{algorithm}
\usepackage{algorithmic}
\usepackage[ansinew]{inputenc}

\newtheorem{cor}{Corollary}
\newtheorem{alg}{Algorithm}
\newcommand{\prob}{{\cal P}}
        \newtheorem{definition}{Definition}%
        \newtheorem{theorem}{Theorem}%
        \newtheorem{lemma}{Lemma}%
\DeclareOldFontCommand{\rm}{\normalfont\rmfamily}{\mathrm}
\DeclareOldFontCommand{\sf}{\normalfont\sffamily}{\mathsf}
\DeclareOldFontCommand{\tt}{\normalfont\ttfamily}{\mathtt}
\DeclareOldFontCommand{\bf}{\normalfont\bfseries}{\mathbf}
\DeclareOldFontCommand{\it}{\normalfont\itshape}{\mathit}
\DeclareOldFontCommand{\sl}{\normalfont\slshape}{\@nomath\sl}
\DeclareOldFontCommand{\sc}{\normalfont\scshape}{\@nomath\sc}

\newcommand{\comment}[1]{}

\newcommand{\bml}[1]{\begin{multline}\label{#1}}
\newcommand{\eml}{\end{multline}}
\newcommand{\beq}[1]{\begin{equation}\label{#1}}
\newcommand{\eeq}{\end{equation}}
\newcommand{\beann}{\begin{eqnarray*}}
\newcommand{\eeann}{\end{eqnarray*}}
\newcommand{\bea}[1]{\begin{eqnarray}\label{#1}}
\newcommand{\eea}{\end{eqnarray}}
\newcommand{\bmp}{\begin{minipage}}
\newcommand{\emp}{\end{minipage}}

\newcommand{\definref}[1]{{\itshape Definition~\ref{#1}}}
\newcommand{\theorref}[1]{{\itshape Theorem~\ref{#1}}}
\newcommand{\lemmaref}[1]{{\itshape Lemma~\ref{#1}}}
\newcommand{\corolref}[1]{{\itshape Corollary~\ref{#1}}}

\newcommand{\secref}[1]{Section~\ref{#1}}

\newcommand{\fig}[1]{{Fig.~\ref{#1}}}

\def\SET0N {I\hspace{-0.8ex}N_0}

{%
}%

{%
}%
\newsavebox{\Citname}

\newcommand{\ignore}[1]{}

\def\NN{$\mathbb{N}$}

\begin{document}

\title{Network Information Flow in\\ Small-World Networks}
\author{Rui A. Costa \hspace{1cm} Jo\~ao Barros
\thanks{The authors are with the Laboratory of Artificial Intelligence and
Computer Science (LIACC/UP) and the Department of Computer Science of the School
of Sciences of the University of Porto, Porto, Portugal. URL: {\tt
http://www.dcc.fc.up.pt/\~{ }barros/}. Work partly supported by the
Funda\c{c}\~ao para
a Ci\^encia e Tecnologia (Portuguese Foundation for Science and Technology)
under grant POSC/EIA/62199/2004. Parts of this work have been presented at the
2006 IEEE Workshop on Information Theory held in Punta del Este, Uruguay,
~\cite{Costa-Barros:06a}, at the 2006 IEEE Workshop on Network Coding, Theory
and Applications,~\cite{Costa-Barros:06b}.}  }
\date{}

\maketitle

\begin{abstract}
Recent results from statistical physics show that large classes of complex
networks, both man-made and of natural origin, are characterized by high
clustering properties yet strikingly short path lengths between pairs of nodes.
This class of networks are said to have a small-world topology.
In the context of communication networks, navigable small-world
topologies, i.e. those which admit efficient distributed routing algorithms, are
deemed particularly effective, for example in resource discovery tasks and
peer-to-peer applications. Breaking with the traditional
approach to small-world topologies that privileges graph parameters pertaining
to connectivity, and intrigued by the fundamental limits of communication in
networks that exploit this type of topology, we investigate the 
capacity of these networks from the perspective of network information
flow. Our contribution includes upper and lower bounds for the capacity of
standard and navigable small-world models, and the somewhat surprising result
that, with high probability, random rewiring does not alter the capacity of a
small-world network.
\end{abstract}

\newpage
\section{Introduction}
\label{sec:intro}

\subsection{Small-World Graphs}

Small-world graphs, i.e. graphs with high clustering coefficients and small
average path length, have sparked a fair amount of interest from the scientific
community, mainly due to their ability to capture fundamental properties of
relevant phenomena and structures in sociology, biology, statistical physics and
man-made networks. Beyond well-known examples such as Milgram's ''six degrees of
separation"~\cite{milgram:67} between any two people in the United States (over
which some doubt has recently been casted~\cite{kleinfeld:02}) and the Hollywood
graph with links between actors, small-world structures appear in such diverse
networks as the U.S. electric power grid, the nervous system of the nematode
worm {\it Caenorhabditis elegans}~\cite{yamamoto:92}, food
webs~\cite{martinez:00}, telephone call graphs~\cite{westbrook:98}, citation
networks of scientists~\cite{newman:01}, and, most strikingly,  the World Wide
Web~\cite{broder:00}.

\begin{figure}[h!]
    \centering
        \includegraphics[width=4cm]{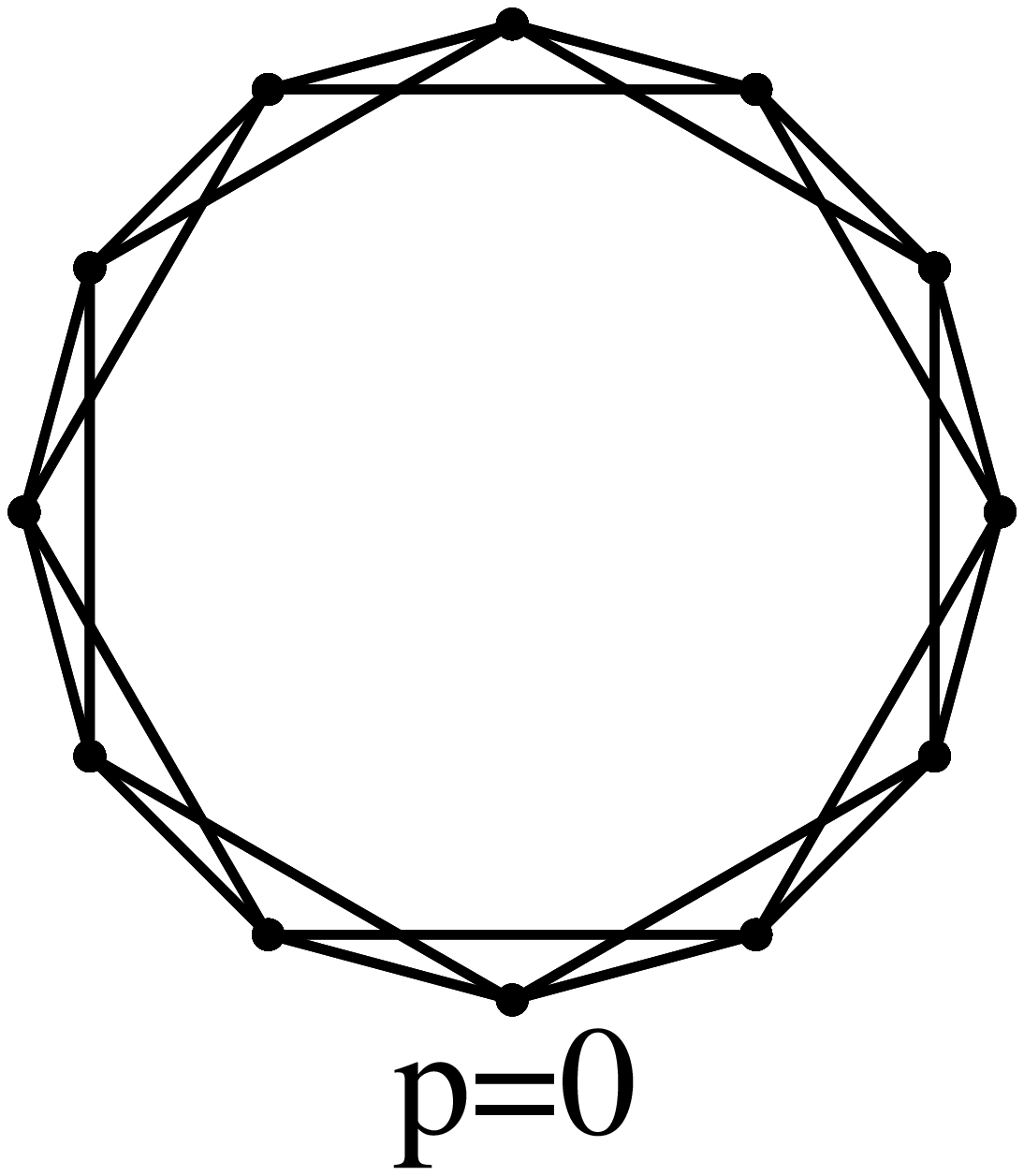}\hspace{0.35cm}
        \includegraphics[width=4cm]{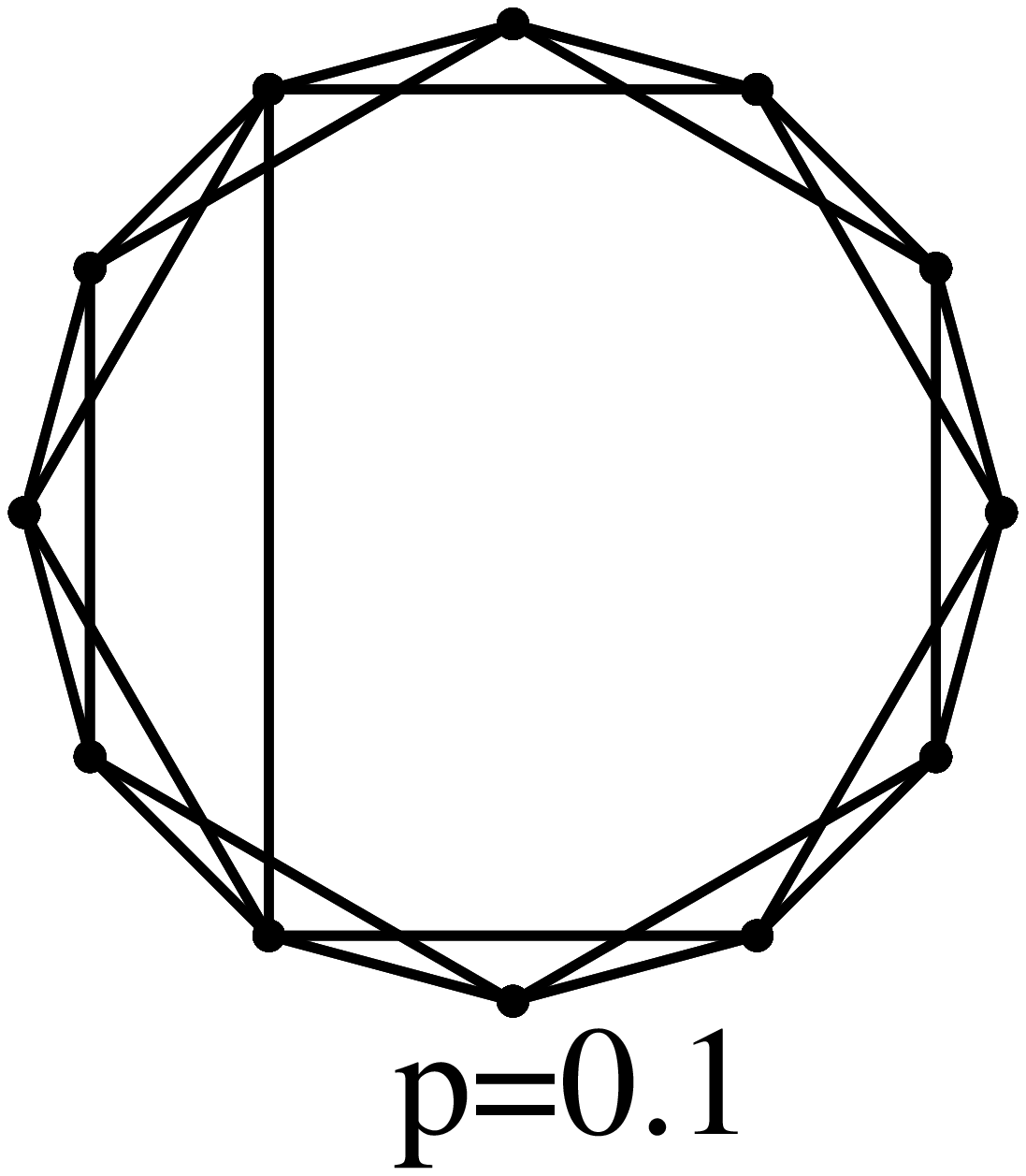} \hspace{0.25cm}
        \includegraphics[width=4cm]{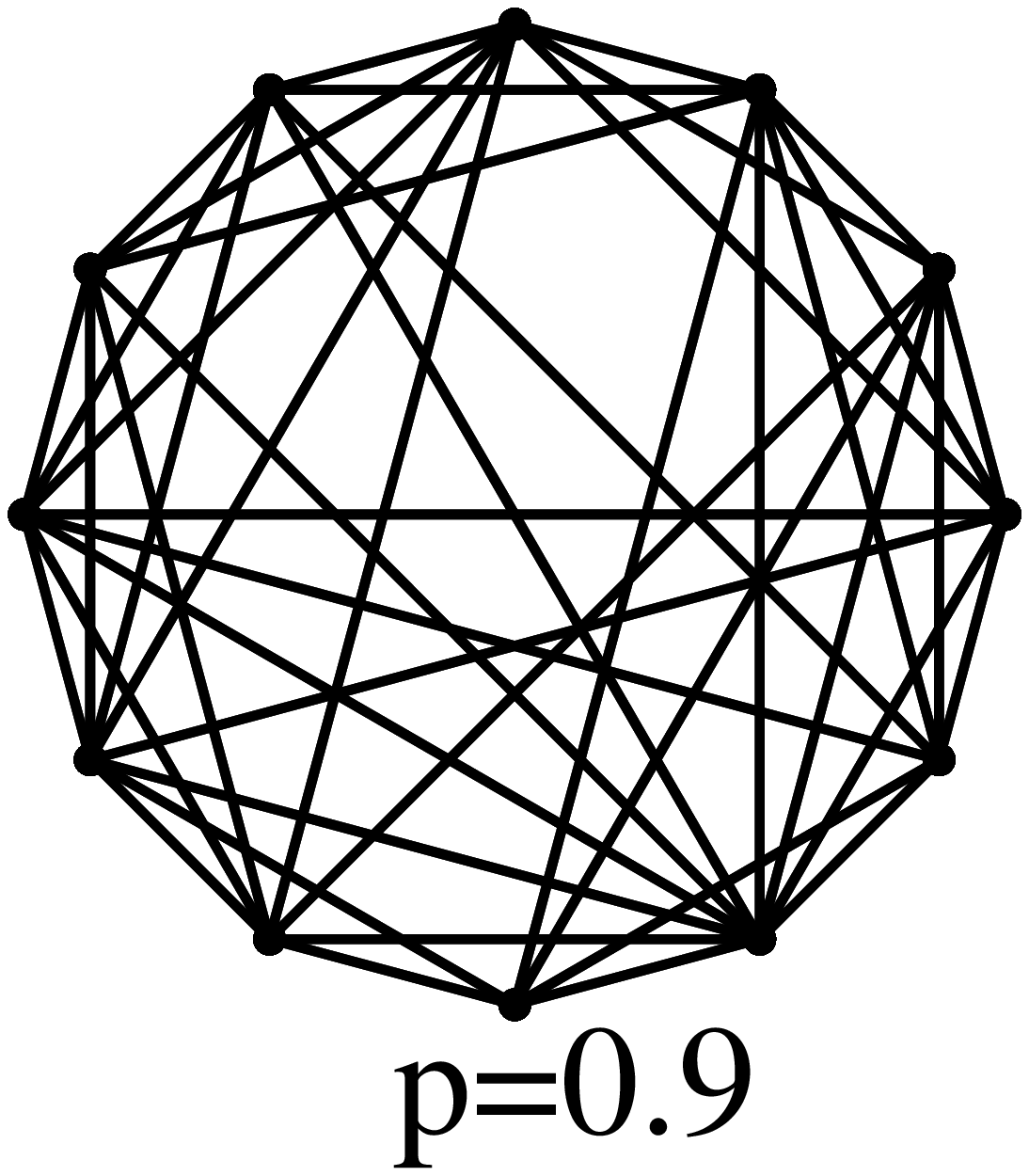}
    \caption{Small-world model with added shortcuts for different values of the
adding probability $p$.}
    \label{fig:interpolation1}
\end{figure}

\begin{figure}[h!]
    \centering
        \includegraphics[width=4cm]{k4p0.eps}\hspace{0.35cm}
        \includegraphics[width=4cm]{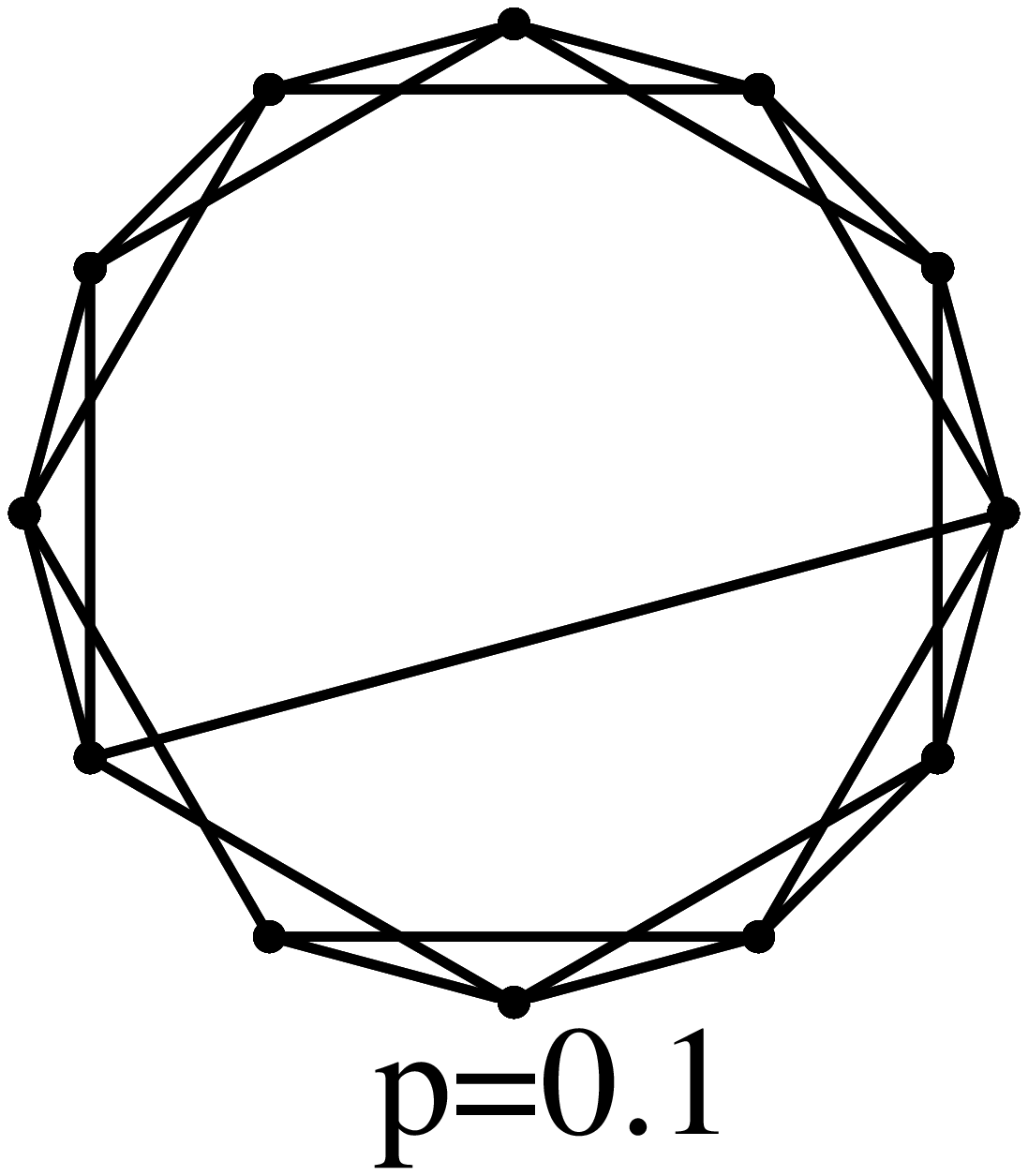} \hspace{0.25cm}
        \includegraphics[width=4cm]{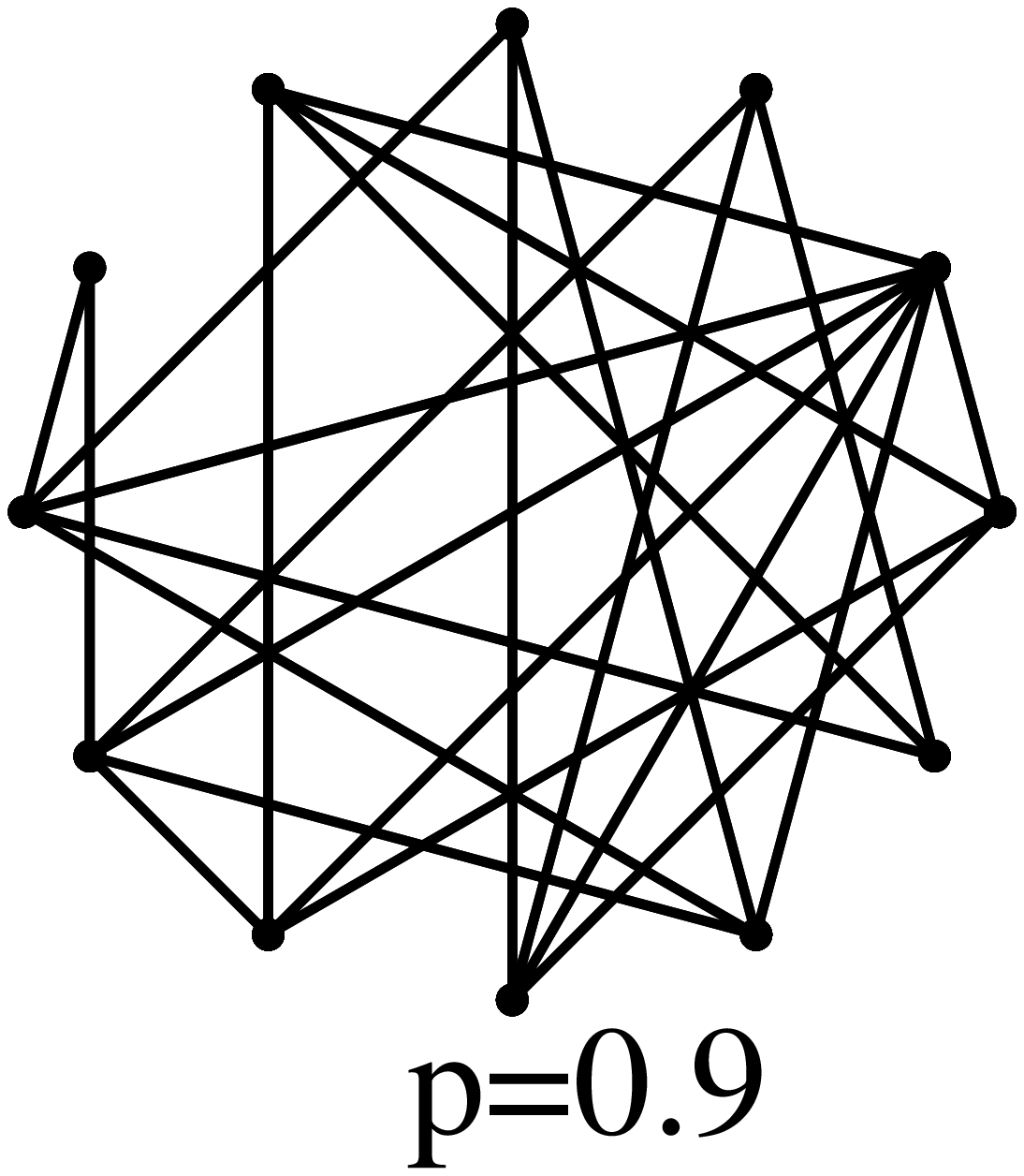}
    \caption{Small-world model with rewiring for different values of the
rewiring probability $p$.}
    \label{fig:interpolation2}
\end{figure}

The term small-world graph itself was coined by Watts and Strogatz,
who in their seminal paper ~\cite{watts:98} defined a class of
models which interpolate between regular lattices and random
Erd\"os-R\'enyi graphs by adding shortcuts or rewiring edges with a
certain probability $p$ (see Figures \ref{fig:interpolation1} and
\ref{fig:interpolation2}). The most striking feature of these models
is that for increasing values of $p$ the average shortest-path
length diminishes sharply, whereas the clustering coefficient
remains practically constant during this transition.


%The combination of strong local connectivity and long-range shortcut
%links renders small-world topologies potentially attractive in the
%context of communication networks, either to increase capacity or
%simplify certain tasks. Recent examples include resource discovery
%in wireless networks~\cite{helmy:03} and design of heterogeneous
%networks~\cite{reznik:03, Dixit:05}. Another relevant application is
%related to overlay networks for peer-to-peer~communications, for
%which small world properties are deemed to be particularly
%useful~\cite{Manku:04}.

Since small-world graphs were proposed as models for complex
networks~\cite{watts:98} and ~\cite{NewmanW:99}, most contributions
in the area of complex networks focus essentially on connectivity
parameters such as the degree distribution, the clustering
coefficient or the shortest path length between two nodes (see
e.g.~\cite{strogatz:01}) . In spite of its arguable relevance ---
particularly where communication networks are concerned --- the {\it
capacity} of small-world networks has, to the best of our knowledge,
not yet been studied in any depth by the scientific community. The
main goal of this paper is thus to provide a preliminary
characterization of the capacity of small-world networks from the
point of view of network information flow.

\subsection{Related Work}
\label{sec:related}

Although the capacity of networks (described by general weighted graphs)
supporting multiple communicating parties is largely unknown, progress has
recently been reported in
several relevant instances of this problem. In the case where the
network has one or more independent sources of information but only
one sink, it is known that routing offers an optimal solution for
transporting messages~\cite{Lehman:04} --- in this case the
transmitted information behaves like {\it water in pipes} and the
capacity can be obtained by classical network flow methods.
Specifically, the capacity of the network follows from the
well-known Ford-Fulkerson {\it max-flow min-cut}
theorem~\cite{ford:62}, which asserts that the maximal amount of a
flow (provided by the network) is equal to the capacity of a minimal
cut, i.e.~a nontrivial partition of the graph node set $V$ into two
parts such that the sum of the capacities of the edges connecting
the two parts (the cut capacity) is minimum. In~\cite{BarrosS:06} it
was shown that network flow methods also yield the capacity for
networks with multiple {\it correlated} sources and one sink.

The case of general multicast networks, in which a single source
broadcasts a number of messages to a set of sinks, is considered
in~\cite{AhlswedeCLY:00}, where it is shown that applying coding
operations at intermediate nodes (i.e.~{\it network coding}) is necessary to
achieve the max-flow/min-cut bound of the network. In
other words, if $k$ messages are to be sent then the minimum cut
between the source and each sink must be of size at least $k$. A
converse proof for this problem, known as the {\it network
information flow problem}, was provided by~\cite{Borade:02}, whereas
linear network codes were proposed and discussed in~\cite{LiYC:03}
and~\cite{KoetterM:03}. Max-flow min-cut capacity bounds for
Erd\"os-R\'enyi graphs and random geometric graphs were presented
in~\cite{ramamoorthy:03}.

Another problem in which network flow techniques have been found
useful is that of finding the maximum stable throughput in certain
networks.  In this problem, posed by Gupta and Kumar
in~\cite{GuptaK:00}, it is sought to determine the maximum rate at
which nodes can inject bits into a network, while keeping the system
stable.  This problem was reformulated in~\cite{PerakiS:05} as a
multi-commodity flow problem, for which tight bounds were obtained
using elementary counting techniques.

Since the seminal work of~\cite{watts:98}, key properties of
small-world networks, such as clustering coefficient, characteristic
path length, and node degree distribution, have been studied by
several authors (see e.g.~\cite{mendes:03} and references therein).
The combination of strong local connectivity and long-range shortcut
links renders small-world topologies potentially attractive in the
context of communication networks, either to increase their capacity
or simplify certain tasks. Recent examples include resource
discovery in wireless networks~\cite{helmy:03}, design of
heterogeneous networks ~\cite{reznik:03, Dixit:05}, and peer-to-peer
communications~\cite{Manku:04}.

When applying small-world principles to communication networks, we
would like not only that short paths exist between any pairs of
nodes, but also that such paths can easily be found using merely
local information. In ~\cite{kleinberg:00} it was shown that this
{\it navigability} property, which is key to the existence of
effective distributed routing algorithms, is lacking in the
small-world models of ~\cite{watts:98} and ~\cite{NewmanW:99}. The
alternative navigable model presented in~\cite{kleinberg:00}
consists of a grid to which shortcuts are added not uniformly but
according to a harmonic distribution, such that the number of
outgoing links per node is fixed and the link probability depends on
the distance between the nodes. For this class of small-world
networks a {\it greedy} routing algorithm, in which a message is
sent through the outgoing link that takes it closest to the
destination, was shown to be effective, thus opening the door
towards a capacity-attaining solution.

\subsection{Our Contributions}

We provide a set of upper and lower bounds for the max-flow min-cut
capacity of several classes of small-world networks, including
navigable topologies, for which highly efficient distributed routing
algorithms are known to exist and distributed network coding
strategies are likely to be found. Our main contributions are as
follows:

\begin{itemize}

\item {\it Capacity Bounds on Small-World Networks with Added Shortcuts:}
We prove a high concentration result which gives upper and lower bounds on the
capacity of a small-world with shortcuts of probability $p$, thus describing
the capacity growth due to the addition of random edges.

\item {\it Rewiring does not alter the Capacity:} We construct assymptotically
tight upper and lower bounds for the capacity of small-worlds with rewiring and
prove that, with high probability, capacity  will not change when the edges are
altered in a random fashion.

\item {\it Capacity Bounds for Kleinberg Networks:} We construct upper and lower
bounds for
the max-flow min-cut capacity of navigable small-world networks  derived from a
square lattice and illustrate how the
choice of connectivity parameters affects communication.

\item {\it  Capacity Bounds for  Navigable Small-World Networks on Ring
Lattices:}
Arguing that the corners present in the aforementioned Kleinberg networks
introduce undesirable artefacts in the computation of the capacity, we define a
navigable small world network based on a ring lattice, prove its navigability
and derive a high-concentration result for the capacity of this instance, as
well.
\end{itemize}

The rest of the paper is organized as follows.
\secref{sec:swm} establishes some notation and offers precise definitions for
the small-world models of interest in this work. Our main results are stated
and proved in Sections III and IV, for classical and navigable networks,
respectively. The paper concludes with \secref{sec:conclusions}.

\section{Classes of Small-World Networks}
\label{sec:swm}

\begin{figure}[h!]
     \centering
        \includegraphics[width=5cm]{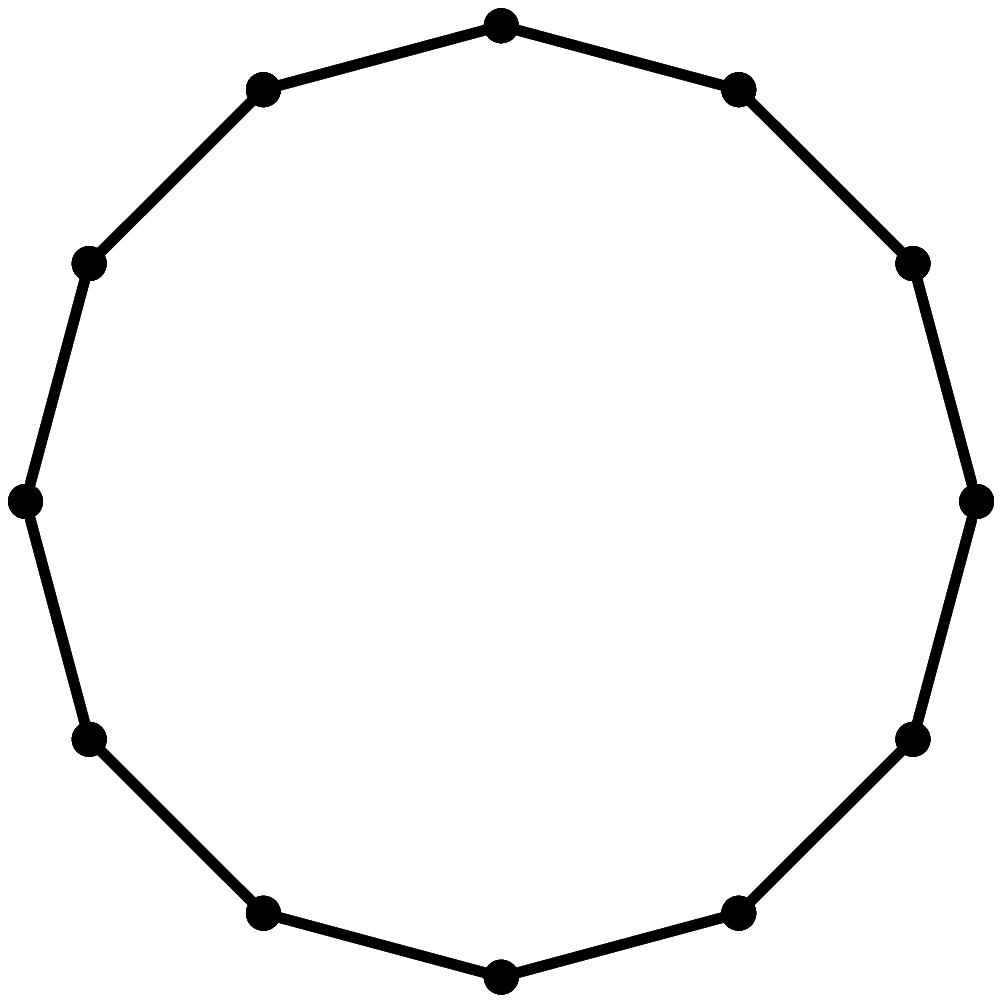}\hspace{-0.45cm}
        \includegraphics[width=5cm]{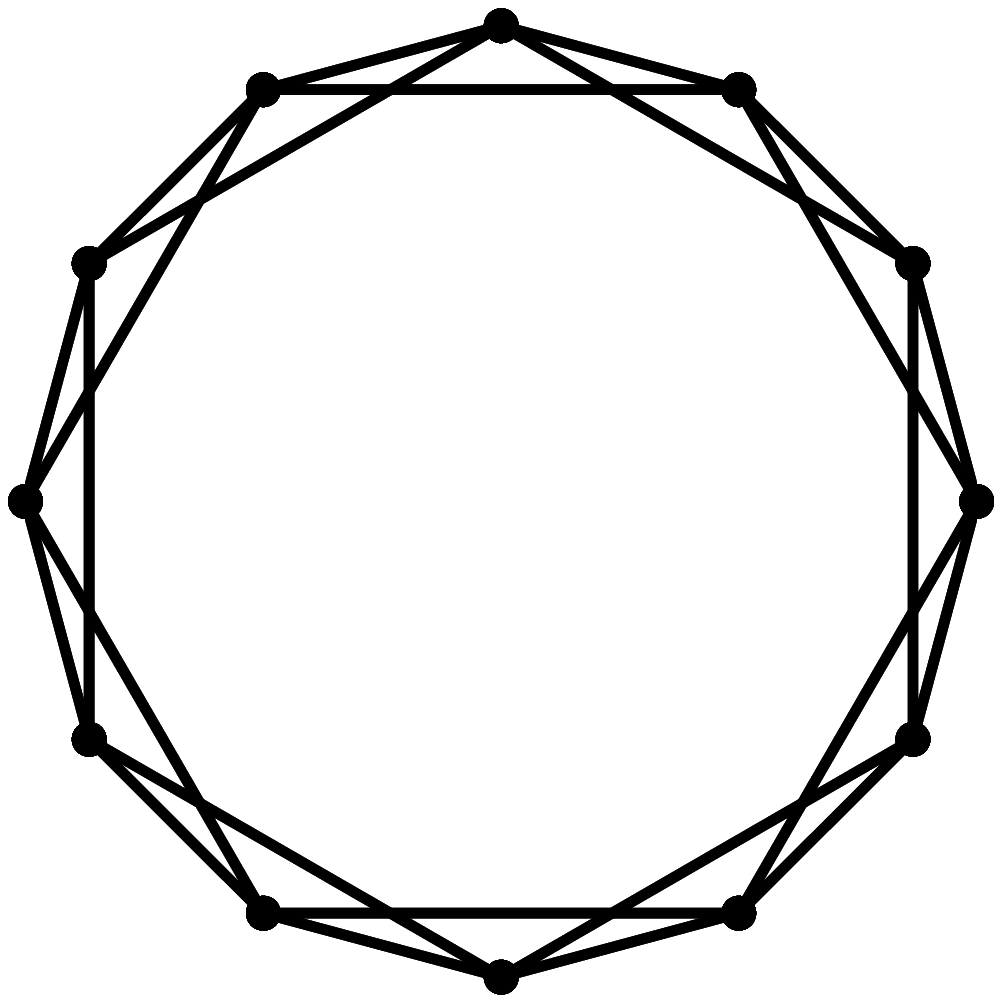} \hspace{-0.45cm}
        \includegraphics[width=5cm]{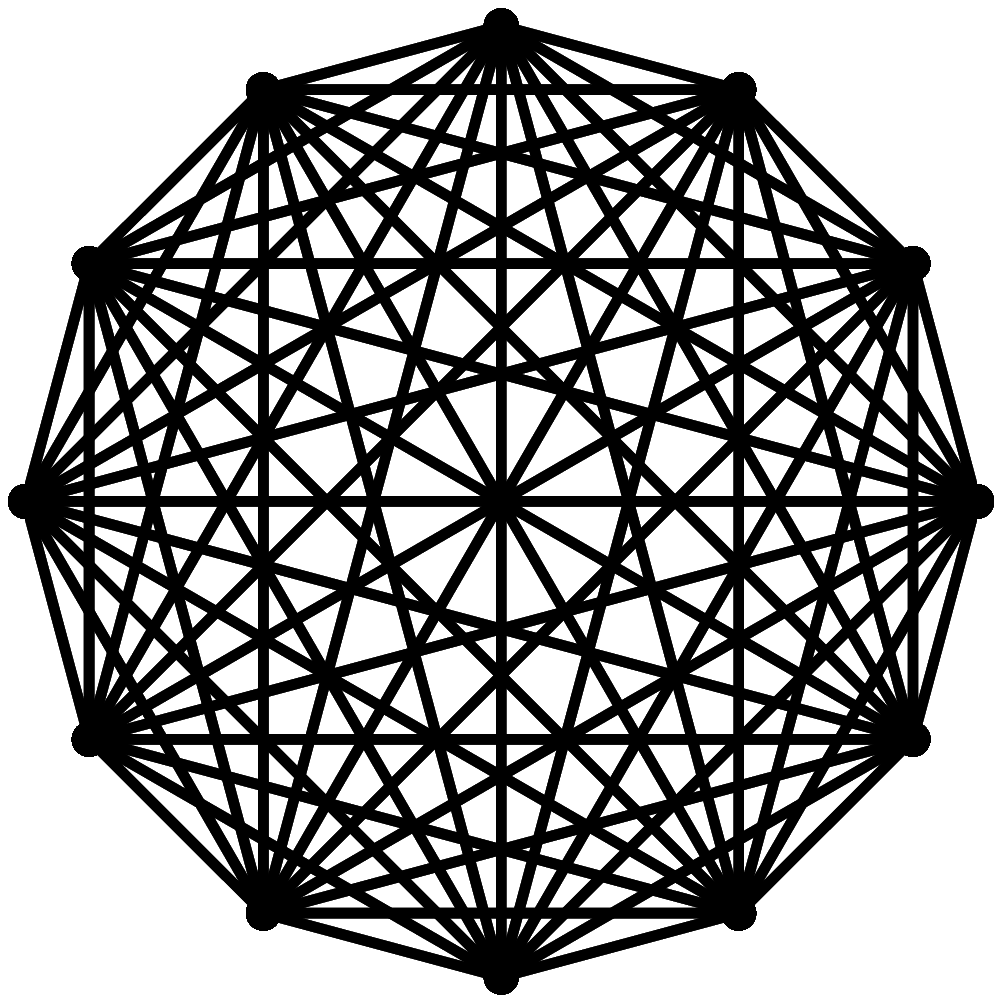}
    \caption{Illustration of a k-connected ring lattice: from left to right
    $k=2, 4, 12$.}
    \label{fig:ringlattice}
\end{figure}

In this section, we give rigorous definitions for the
classes of small-world networks under consideration.
First, we require a precise notion of
distance in a ring.
\begin{definition}
Consider a set of $n$ nodes connected by edges that form a ring
(see \fig{fig:ringlattice}, left plot). The {\it ring distance}
between two nodes is defined as the minimum number of {\it hops}
from one node to the other. If we number the nodes in clockwise
direction, starting from any node, then the ring distance between
nodes $i$ and $j$ is given by $d(i,j)=min\{|i-j|,n+i-j,n-|i-j|\}.$
\end{definition}
For simplicity, we refer to $d(i,j)$ as the {\it distance} between
$i$ and $j$. Next, we define a $k$-connected ring lattice.

The ring lattice that serves as basis for some of the small-world
models described next, can be defined as follows.
\begin{definition}
A $k$-{\it connected ring lattice} (see \fig{fig:ringlattice}) is a
graph $L=(V_L,E_L)$ with nodes $V_L$ and edges $E_L$,
in which all nodes in $V_L$ are placed on a ring and are connected
to all the nodes within distance $\frac{k}{2}$.
\end{definition}

Notice that in a $k$-connected ring lattice, all the nodes have degree $k$.
We are now ready to define the targeted small-world models.

\begin{definition}[Small-World Network with Shortcuts~\cite{NewmanW:99},
see~\fig{fig:interpolation1}]
\label{def:shortcuts}
Consider a $k$-connected ring lattice $L=(V_L,E_L)$ and let $E_C$ be the set of
all possible edges between nodes in $V_L$. To obtain a
{\it small-world network with shortcuts}, we add
to the ring lattice $L$ each edge $e\in E_C \backslash E_L$
with probability $p$.
\end{definition}

\begin{definition}[Small-World Network with Rewiring~\cite{watts:98},
see~\fig{fig:interpolation2}]
\label{def:watstro} To obtain a {\it small-world network with rewiring}, we use
the following procedure. Consider a
$k$-connected ring lattice $L=(V_L,E_L)$ and choose a node, say node $u$, and
the edge that connects it to its nearest neighbor in a clockwise sense. With
probability $p$, reconnect this edge to a node chosen uniformly at random over
the set of nodes $\left\{ v\in V_L:(u,v)\notin E_L \right\}$. Repeat this
process by moving around the ring in clockwise direction, considering each node
in turn until one lap is completed. Next, consider the edges that connect nodes
to their second-nearest neighbors clockwise. As before, randomly rewire each of
these edges with probability $p$, and continue this process, circulating around
the ring and proceeding outward to more distant neighbors after each lap, until
each edge in the original lattice has been considered once.
\end{definition}

\begin{figure}[h!]
    \centering
        \includegraphics[width=7cm]{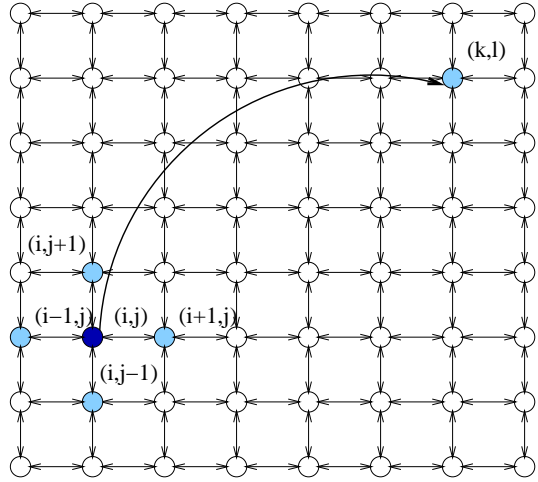}
    \caption{Kleinberg network. Each node is directly connected to all neighbors
within $h$ hops, and also to $q$ more distant nodes through so called shortcuts.
In the shown example, where $h=q=1$, lightly shaded circles represent the nodes
that are directly connected to node $(i,j)$, i.e.~the four direct neighbors of
$(i,j)$ and one additional node $(k,l)$ connected by a shortcut.}
    \label{fig:kleinberg}
\end{figure}

\begin{definition}[Kleinberg Network~\cite{kleinberg:00},
see~\fig{fig:kleinberg}]
\label{def:klein} We begin from a two-dimensional grid and a set of nodes that
are identified with the set of lattice points in an $n\times n$ square,
$\{(x,y): x\in \{ 1,2,...,n\},y\in \{ 1,2,...,n\} \}$, and we define the {\it
lattice distance} between two nodes $(x_1,x_2)$ and $(y_1,y_2)$ to be the number
of {\it lattice steps} (or hops) separating them: $d(x,y)=|y_1-x_1|+|y_2-x_2|$.
For a constant $h\geq 1$, the node $(u_1,u_2)$ ($\forall u_1,u_2\in \left\{
1,...,n \right\}$) is connected to every other node within lattice distance $h$
(we denote the set of this initial edges as $E_L$). For universal constants
$q\geq 0$ and $r\geq 0$, we also construct edges from $u$ to $q$ other nodes
using random trials; the $i^{th}$ edge from $u$ has endpoint $v$ with
probability proportional to $d(u,v)^{-r}$. To ensure a valid probability
distribution, consider the set of nodes that are not connected with $u$ in the
initial lattice, $N_u=\{w:d(u,w)>h\}$, and divide $d(u,v)^{-r}$ by the
appropriate normalizing constant $s(u)=\sum_{w\in N_u}
[d(u,w)]^{-r}.$
\end{definition}

In the next section, we will see that this model exhibits unexpected effects
related to the corners of the chosen base lattice. Motivated by this
observation, we construct a somewhat different model, which uses a ring lattice
but still keeps the key relationship between shortcut probability and node
distance that assures the navigability of the model (as proven in the appendix).

\begin{definition}[Navigable Small-World Ring]
\label{def:navi} Consider a $k$-connected ring lattice. For universal constants
$q \geq 0$ and $r \geq 0$, we add new edges from node $i$ ($\forall i$) to $q$
other nodes randomly: each added edge has an endpoint $j$ with probability
proportional to $d(i,j)^{-r}$. To ensure a valid probability distribution,
consider $N_i=\left\{ j: d(i,j)>\frac{k}{2} \right\}$ and divide $d(i,j)^{-r}$
by the appropriate normalizing constant $s_i=\sum_{j\in N_i} d(i,j)^{-r}$.
\end{definition}

\section{Capacity Results for Small-World Networks}
\label{sec:cap}

In \secref{sec:related}, we argued that the max-flow min-cut capacity provides
the fundamental limit of communication for various relevant network scenarios.
Motivated by this observation, we will now use network flow methods and random
sampling techniques in graphs to derive a set of bounds for the capacity of the
small-world network models presented in the previous section. Although all of
the models discussed in this section are based on ring lattices, it is worth
pointing out that the methodology presented next can be equally applied to other
classes of base lattices.

\subsection{Preliminaries}

We start by introducing some necessary mathematical tools. Let $G$ be an
undirected graph, representing a communication network, with edges of unitary
weight.
 In the spirit of the max-flow min-cut theorem of Ford and
Fulkerson~\cite{ford:62}, we will refer to the global minimum cut of $G$ as the
max-flow min-cut capacity (or simply the {\it capacity}) of the graph.

Let $G_s$ be the graph obtained by sampling on $G$, such that each edge $e$ has
sampling probability $p_{e}$. From $G$ and $G_s$, we obtain $G_w$ by assigning
to each edge $e$ the weight $p_{e}$, i.e.~$w(e)=p_{e},\forall e$. We denote the
capacity of $G_s$ and $G_w$ by $c_s$ and $c_w$, respectively. It is helpful to
view a cut in $G_s$ as a sum of Bernoulli experiences, whose outcome determines
if an edge $e$ connecting the two sides of the cut belongs to $G_s$ or not. It
is not difficult to see that the value of a cut in $G_w$ is the expected value
of the same cut in $G_s$.
The next theorem gives a characterization of how close a cut in $G_s$ will be
with respect to its expected value.

\begin{theorem}[From~\cite{karger:94}]
Let $\epsilon=\sqrt{2(d+2)\ln(n)/c_w}$. Then, with probability
$1-O(1/n^{d})$, every cut in $G_s$ has value between $(1-\epsilon)$ and
$(1+\epsilon)$ times its expected value.
\label{th:karger}
\end{theorem}

Notice that although $d$ is a free parameter, there is a strict relationship
between the value of $d$ and the value of $\epsilon$. In other words, the
proximity to the expected value of the cut is intertwined with how close the
probability is to one. \theorref{th:karger} yields also the following useful
property.

\begin{cor}
Let $\epsilon=\sqrt{2(d+2)\ln(n)/c_w}$. Then, with high probability,
the value of $c_s$ lies between $(1-\epsilon)c_w$ and $(1+\epsilon)c_w$.
\label{cor:karger}
\end{cor}
\vspace{-0.21cm}
Before using the previous random sampling results to determine bounds for the
capacities of small-world models, we prove another useful lemma.

\begin{lemma}
Let $L=(V_L, E_L)$ be a $k$-connected ring lattice  and  let $G=(V_L,E)$ be a
fully connected graph (without self-lops), in which edges $e\in E_L$ have weight
$w_1\geq 0$ and edges $f\notin E_L$ have weight $w_2\geq 0$. Then, the global
minimum cut in $G$ is $k w_1+(n-1-k)w_2$.
\label{lemma:mincut}
\end{lemma}

\begin{proof}
We start by splitting $G$ into two subgraphs: a $k$-connected ring lattice $L$
with weights $w_1$ and a graph $F$ with nodes $V_L$ and all remaining edges of
weight $w_2$. Clearly, the value of a cut in $G$ is the sum of the values of the
same cut in $L$ and in $F$. Moreover, both in $L$ and in $F$, the global minimum
cut is a cut in which one of the partitions consists of one node (any other
partition increases the number of outgoing edges). Since each node in $L$ has
$k$ edges of weight $w_1$ and each node in $F$ has the remaining $n-1-k$ edges
of weight $w_2$, the result follows.
\end{proof}

\subsection{Capacity Bounds for Small-World Networks with Added Shortcuts}

With this set of tools, we are ready to state and prove our first main result.

\begin{theorem}
With high probability, the value of the capacity of a small-world network with
added shortcuts lies between $(1-\epsilon)c_w$ and $(1+\epsilon)c_w$, with
$\epsilon=\sqrt{2(d+2)\ln(n)/c_w}$ and $c_w=k+(n-1-k)p.$
\end{theorem}

\begin{proof}
Let $G_w$ be a fully connected graph with $n$ nodes and with the edge weights
(or equivalently, the sampling probabilities) defined as follows:
\begin{itemize}
\item The weight of the edges in the initial lattice of a small-world network
with added shortcuts is one (because they are not removed);
\item The weight of the remaining edges is $p$, (i.e.~the probability that an
edge is added).
\end{itemize}
Notice that $G_w$ is a graph in the conditions of \lemmaref{lemma:mincut}, with
$w_1=1$ and $w_2=p$. Therefore, the global minimum cut in $G_w$ is
$c_w=k+(n-1-k)p$, where $k$ is the initial number of neighbors in the lattice.
Using \corolref{cor:karger}, the result follows.
\end{proof}

\begin{figure}[h!]
\begin{center}
\includegraphics[width=10cm]{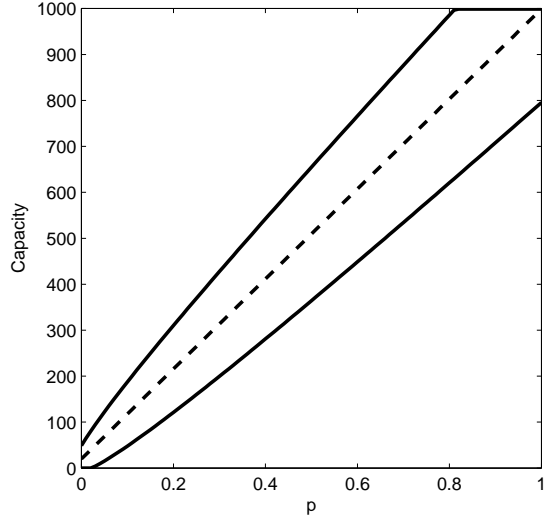}\vspace{-0.3cm}
\caption{Bounds on the capacity of a small-world network with added shortcuts,
for $n=1000$, $k=20$, and $d=1$. The dashed line represents the expected value
of the capacity, and the solid lines represent the bounds. Naturally, the
capacity increases with $p$, as the number of added links become larger.}
\label{fig:addingbounds}
\end{center}
\end{figure}

The obtained bounds are illustrated in \fig{fig:addingbounds}.

\subsection{Capacity Bounds for Small-World Networks with Rewiring}

In the previous classes of small-world networks, edges were added to a
$k$-connected ring lattice (with minimum cut $k$) and clearly the capacity could
only grow with $p$. The next natural step is to ask what happens when edges are
not added but rewired with probability $p$, as described in \secref{sec:swm}.
Before presenting a theorem that answers this question, we will prove the
following lemma.

\begin{lemma}
Let $G_w$ be a weighted, fully connected graph, whose weights correspond to the
edge probabilities of a small-world network with rewiring, and let $c_w$ be the
global minimum cut in $G_w$. Then, $c_w\geq k$.
\label{lemma:gw}
\end{lemma}

\begin{proof}
We start with the initial lattice edges $(l,m)\in E_L$, and assign the weight
$1-p$ to their counterparts in $G_w$. In order to determine the weight of the
non-initial edges that result from rewiring, consider the following events:
\begin{itemize}
\item $R(i,j)$: ``Choose the edge $(i,j)\in E_L$ to be rewired";
\item $C_i(j,l)$: ``Rewire $(i,j)\in E_L$ to $(i,l)\notin E_L$".
\end{itemize}
Notice that $\prob(R(i,j))=p, \forall (i,j)\in E_L$. \\
Let $i$ and $j$ be two non-initially connected nodes. The notation
$i\leftrightarrow j$ denotes the event that the nodes $i$ and $j$ are connected.
We have that
\begin{eqnarray*}
\prob(i\leftrightarrow j)&=&\prob \left([\cup_{x=1}^{k/2} (R(i,i+x)\cap
C_i(i+x,j))] \cup [\cup_{x=1}^{k/2}
(R(j,j+x)\cap C_j(j+x,i))]\right)\\
&=& \prob \left( \cup_{x=1}^{k/2} (R(i,i+x)\cap C_i(i+x,j)) ) +\prob(
\cup_{x=1}^{k/2} (R(j,j+x)\cap C_j(j+x,i)) \right)\\&&-\prob \left(
[\cup_{x=1}^{k/2} (R(i,i+x)\cap C_i(i+x,j))] \cap
[\cup_{x=1}^{k/2} (R(j,j+x)\cap C_j(j+x,i))] \right)
\end{eqnarray*}
Because we do not consider multiple edges, we have that the events $R(i,i+x)\cap
C_i(i+x,j)$ and $R(j,j+y)\cap C_j(j+y,i)$ are mutually exclusive, $\forall x,y$.
Therefore,
\begin{eqnarray*}
\prob(i\leftrightarrow j)&=& \sum_{x=1}^{k/2} \left( \prob(R(i,i+x)\cap
C_i(i+x,j)) +\prob(R(j,j+x)\cap
C_j(j+x,i))\right)\\
&=& \sum_{x=1}^{k/2} \left(\prob(C_i(i+x,j)|R(i,i+x))\prob(R(i,i+x))
+\prob(C_j(j+x,i)|R(j,j+x))\prob(R(j,j+x)) \right)\\
&=& p\cdot \left(\sum_{x=1}^{k/2}
(\prob(C_i(i+x,j)|R(i,i+x))+\prob(C_j(j+x,i)|R(j,j+x)))\right).
\end{eqnarray*}
We have $\prob(C_i(i+x,j)|R(i,i+x))=\frac{1}{m}$, where $m$ is the number of
possible new connections from node $i$ when we rewired the edge $(i,i+x)$. It is
possible that, occurring some rewiring or not, none of the choices to a new link
is the node $i$. In this case, $m=n-k-1$. Notice that this is the highest it can
get, therefore $m\leq n-k-1$. Then, we have
$$\prob(C_i(i+x,j)|R(i,i+x))\geq
\frac{1}{n-k-1}.$$ Analogously, $\prob(C_j(j+x,i)|R(j,j+x)))\geq
\frac{1}{n-k-1}$. Therefore, $$\prob(i\leftrightarrow j)\geq p\cdot \left(
\sum_{x=1}^{k/2} \frac{2}{n-k-1} \right)=\frac{pk}{n-k-1}.$$
There are $k$ initial edges and $n-k-1$ non-initial edges in each node.

Consider a fully connected, weighted graph $F$ with the weights defined as
follows: all the edges $(i,j)\notin E_L$ have the weight $\frac{pk}{n-k-1}$, and
all the others edges $(i,j)\in E_L$ have the weight $1-p$. Notice that $F$ is a
graph in the conditions of Lemma $1$, with $w_1=1-p$ and $w_2=\frac{pk}{n-k-1}$.
Therefore, $$c_F =k(1-p)+(n-k-1)\frac{pk}{n-k-1}=k.$$ Notice that, in this
situation, all the weights in $F$ are a lower bound of the weights in $G_w$.
Therefore, a cut in $F$ is a lower bound for the corresponding cut in $G_w$.
Then, the global minimum cut in $F$ is a lower bound for all the cuts in $G_w$,
in particulary, for $c_w$: $c_w\geq c_F =k$.
\end{proof}
With this lemma, we are now ready to state and prove our next result.
\begin{theorem}[Rewiring does not alter capacity.]
With high probability, the capacity of a small-world network with rewiring has a
value in the interval $[(1-\epsilon)k,k]$ with
$\epsilon=\sqrt{2(d+2)ln(n)/k}$.
\end{theorem}
\begin{proof}
Based on \lemmaref{lemma:gw} and \corolref{cor:karger}, we have that, with high
probability, $c_s \geq (1-\epsilon_w)k$, with
$\epsilon_w=\sqrt{2(d+2)ln(n)/c_w}$. Now, from the fact that $c_w\geq k$, we
have that $\epsilon=\sqrt{2(d+2)ln(n)/k}\geq \epsilon_w$. Then, $(1-\epsilon_w)k
\geq (1-\epsilon)k$, and the first part of the result follows.

Next, we prove by contradiction that $c_s \leq k$. Suppose that the proposition
$c_s>k$ is true. Let $c_i$ be the cut in which one of the partitions consists of
node $i$, $i=1,...,n$. Because $c_s$ is the global minimum cut in $G_s$, we have
that $c_i>k$, $\forall i=1,...,n$. Notice that $c_i$ is the degree of node $i$.
Then, because in the $k$-connected ring lattice all nodes have degree $k$ and
all nodes in $G_s$ have degree greater than $k$ (because $c_i>k,\forall i$), we
have that the number of edges in $G_s$ must be greater than the number of edges
in the $k$-connected ring lattice. But this is clearly absurd, because in the
construction of $G_s$, we do not add new edges to the $k$-connected ring
lattice, we just rewire some of the existent edges.
\end{proof}

\section{Capacity Bounds for Navigable Networks}
\label{sec:navi}

As we argue in \secref{sec:intro}, when considering small-world networks as
communication networks, an important aspect is the ability to find short paths
between any pairs of nodes, using only local information. This property
guarantees that efficient distributed routing algorithms exists. Kleinberg, in
his seminal work \cite{kleinberg:00}, proved that this {\it navigability}
property is lacking in the models of Watts and Strogatz, and introduced a new
model (\definref{def:klein}). Motivated by the relevance
of the navigability property, we present, in this section, the capacity bounds
for Kleinberg Networks and for Navigable Small-World Rings.

\subsection{Capacity Bounds for Kleinberg Networks}

Before proceeding with the bounds for the capacity of Kleinberg networks, we
require an algorithm to calculate the normalizing constants $s(x,y)=\sum
_{(i,j)\in N_{(x,y)}} [d((x,y),(i,j))]^{-r}$ for $x,y\in \{1,...n\}$. For this
purpose, we note that the previous sum can be written as
$$s(x,y)=\sum\limits_{(i,j)\neq (x,y)}
[d((i,j),(x,y))]^{-r} - \sum _{(i,j)\notin N_{(x,y)}}[d((i,j),(x,y))]^{-r}.$$
Clearly, the first term can be easily calculated. Thus,
the challenging task is to present an algorithm that deals with the calculation
of $\sum _{(i,j)\notin N_{(x,y)}} [d((i,j),(x,y))]^{-r}$. The nodes $(i,j)\notin
N_{(x,y)}$ are the nodes initially connected to node $(x,y)$, i.e., the nodes at
a distance $t\leq h$ from node $(x,y)$. It is not difficult to see that the
nodes at a distance $t$ from node $(x,y)$ are the nodes in the square line
formed by the nodes $(x-t,y)$, $(x+t,y)$, $(x,y+t)$ and $(x,y-t)$. Then, we
could just look at nodes in the square formed by the nodes $(x-h,y)$, $(x+h,y)$,
$(x,y+h)$ and $(x,y-h)$ and sum all the corresponding distances to node $(x,y)$.
A corner effect occurs when when this square lies outside the base lattice.
Assume that we start by calculating the distances to the nodes in  line $y+i$,
with $i\geq 0$.

\begin{table}[b!]
\caption{Algorithm for computing normalizing constants}
\label{tab:algo}
\begin{center}
\begin{minipage}[h]{8cm}
\begin{alg}
$_{}$\\
$_{}\ \ z=[0]_{n\times n}$\\
$_{}\ \ for \: i=0:min\{h,n-y\} $\\
$_{}\ \ \ \ \ \ for \: j=0:min\{h-i,n-x\}$\\
$_{}\ \ \ \ \ \ \ \ \ \ z(x+j,y+i)=(i+j)^{-r}$\\
$_{}\ \ \ \ \ \ for \: j=1:min\{h-i,x-1\}$\\
$_{}\ \ \ \ \ \ \ \ \ \ z(x-j,y+i)=(i+j)^{-r} $\\
$_{}\ \ for \: i=1:min\{h,y-1\}$\\
$_{}\ \ \ \ \ \ for \: j=0:min\{h-i,n-x\}$\\
$_{}\ \ \ \ \ \ \ \ \ \ z(x+j,y-i)=(i+j)^{-r}$\\
$_{}\ \ \ \ \ \ for \: j=1:min\{h-i,h-(m_1-i),x-1\} $\\
$_{}\ \ \ \ \ \ \ \ \ \ z(x-j,y-i)=(i+j)^{-r} $\\
$_{}\ \ z(x,y)=0$\\
$_{}\ \ z=\sum_{i=1}^{n} \sum_{j=1}^{n} z(i,j)$\\
$_{}\ \ s(x,y)=\sum_{(i,j)\neq
(x,y)} (|i-x|+|j-y|)^{-r}-z$
\end{alg}
 \end{minipage}
\end{center}
\end{table}

To avoid calculating extra distances (i.e., distances of nodes that are out of
the grid), we must make sure that this line verifies $y+i\leq n$ and also
$y+i\leq h$. For this reason, $i$ must vary according to
$i\in\{0\dots\min\{h,n-y\}\}$. Now, in each line $y+i$, we first look at the
nodes in the right side of $(x,y)$, i.e., we calculate the distances of the
nodes $(x+j,y+i)$, with $j\geq 0$. Now, notice that in the line $y$, we have $h$
points on the right side of $(x,y)$ that are in the square (regardless of
whether they are in the grid). Because the distance is the minimum number of
steps in the grid, we have that in line $y+i$ there are $h-i$ points at the
right side of $(x,y)$ that are inside the square. This way,  $j$ must be vary
according to $j\in\{0\dots\min\{h-i,n-x\}\}$. Now, when looking at the nodes at
the left side (i.e., the nodes $(x-j,y+i)$, with $i\geq 1$), the idea is the
same. The only difference is that, in this case, the variation for $j$ is
$j\in\{1\dots\min\{h-i,x-1\}\}$. Then, we proceed analogously
for the lines below $(x,y)$, i.e., the lines $y-i$, with
$i\in\{1\dots\min\{h,y-1\}\}$. This algorithm is summarized in Table
\ref{tab:algo}. The matrix $z$ is a buffer for the distances, i.e.,
$z(u_1,u_2)=d((x,y),(u_1,u_2))$. We impose $z(x,y)=0$, because
$d((x,y),(x,y))^{-r}$ is also calculated in this procedure.

The following quantities will be instrumental towards characterizing the
capacity:
\begin{eqnarray}
M &=& \max\left\{ \frac{h(h+3)}{2}+q ,
(1-\epsilon)c_w\right\}\nonumber\\
\epsilon &=& \sqrt{2(d+2)\ln(n^2)/c_w}\nonumber\\
c_w &=& \frac{h(h+3)}{2}+\sum_{x=1}^{h+1}
\sum_{y=h+2-x}^{n} f(x,y)+\sum_{x=h+2}^{n} \sum_{y=1}^{n}
f(x,y)\label{eq:cw}\\
f(x,y) &=& q\cdot (
g_{(x,y)}(1,1)+g_{(1,1)}(x,y))\nonumber\\
g_{(x,y)}(a,b) &=& \left(
1-\frac{(x+y-2)^{-r}}{s(a,b)} \right) ^{q-1}\cdot
\frac{(x+y-2)^{-r}}{s(a,b)}\nonumber\\
s(1,1) &=& \sum_{i=h+1}^{n-1} (i+1)\cdot
i^{-r}\nonumber+
\sum_{i=0}^{n-2} (n-1-i)\cdot (n+i)^{-r}.\nonumber
\end{eqnarray}
Recall that $s(x,y)$ can be calculated using {\it Algorithm 1}. The proof of the
capacity will rely heavily on the following lemma.

\begin{lemma}
Let $G_w$ be the weighted graph associated with a Kleinberg network, and $c_w$
be the global minimum cut in $G_w$. Then, for $h<n-1$, $c_w$ is given by
(\ref{eq:cw}).
\label{lemma:mincut2}
\end{lemma}

\begin{proof}
All the edges $e\in E_L$ have weight $1$ (because they are never removed), all
nodes in $G_w$ have degree $n^2 -1$, and the weights of these edges depend only
on the distance between the nodes they connect. Therefore, the global minimum
cut in $G_w$ must be a cut in which one of the partitions consists of a single
node. Because the weight of an edge in $G_w$ decreases with the distance between
the nodes that it connects, the global minimum cut in $G_w$ must be a cut in
which one of the partitions consists of a single node that maximizes the
distance to other nodes. Therefore, because a corner node has more nodes at a
greater distance than the other nodes and has also a smaller number of nodes to
which it is connected, $c_w$ must be a cut in which one of the partitions
consists of a corner node: $(1,1),(1,n),(n,1)$ or $(n,n)$.

Assume, without loss of generality, that $c_w$ is the cut in which one of the
partitions consists of node $(1,1)$. Let $w(u,v)$ be the weight of the edge
connecting the nodes $u$ and $v$. This way, $c_w=\sum_{u\neq (1,1)} w((1,1),u)$.
Now, we must count how many edges connecting node $(1,1)$ are in $E_L$,
therefore, having weight $1$. For this, we define an auxiliary way to numerate
diagonals: $\{(1,1)\}$ is the diagonal $0$, $\{(1,2),(2,1)\}$ is diagonal 1, and
so on (see Figure \ref{fig:squaregrid}).

\begin{figure}[h!]
\begin{center}
\includegraphics[width=7cm]{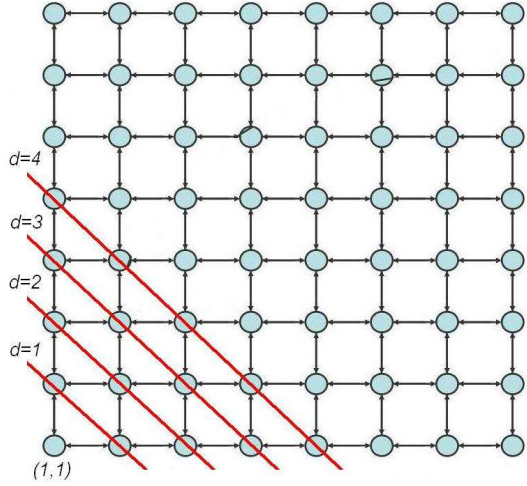}\vspace{-0.3cm}
\caption{Numeration of the diagonals of a square lattice.}
\label{fig:squaregrid}
\end{center}
\end{figure}

It is not difficult to see that the nodes in the $i^{th}$ diagonal have a
distance $i$ to node $(1,1)$, $i\in \{1,...,2(n-1)\}$. Now, for $i\leq n-1$,
there are $i+1$ nodes in the $i^{th}$ diagonal and, for $i=n+j$ with $j\in
\{0,...,n-2\}$, there are $n-1-j$ nodes in the $i^{th}$ diagonal. Then, there
are $\sum_{i=1}^{h} i+1=h(h+3)/2$ nodes initially connected to node $(1,1)$
(again, with $h<n-1$), then there are $h(h+3)/2$ edges with weight 1. Therefore,
we have that: $$c_w=\frac{h(h+3)}{2}+\sum_{x=1}^{h+1} \sum_{y=h+2-x}^{n}
w((1,1),(x,y))+\sum_{x=h+2}^{n}\sum_{y=1}^{n} w((1,1),(x,y)).$$

Next, we determine the weights, $w(u,v)$. Consider two nodes that are not
connected initially, $u=(u_1,u_2)$ and $v=(v_1,v_2)$, and the edge $(u,v)$. This
edge can be added in two different trials: one for node $u$ and another one for
node $v$. Because we do not consider multiple edges, these can be viewed as two
mutually exclusive trials.
Therefore, the weight of this edge is the sum of the
probabilities of adding this edge when considering node $u$ and when considering
node $v$. Let us focus on node $u$. The trial ``add edge $(u,v)$" follows a
Binomial distribution, with $q$ Bernoulli experiences, with success probability
$$a_u(v)=\frac{d(u,v)^{-r}}{s(u)}=\frac{(|u_1-v_1|+|u_2-v_2|)^{-r}}{s(u)}.$$
Therefore, the probability of adding the edge $(u,v)$, when considering node
$u$, is $q\cdot (1-a_u (v))^{q-1}\cdot a_u (v)$.
Therefore, the weight of the edge $((u_1,u_2),(v_1,v_2))$ is
$$w(u,v)=q\cdot (1-a_u (v))^{q-1}\cdot a_u (v)+q\cdot(1-a_v (u))^{q-1}\cdot a_v
(u).$$
As we have seen, the global minimum cut in $G_w$ is the cut in which
one of the partitions consists of node $(1,1)$. We have that, if
$(x,y)$ is a node of the grid, $x\geq 1$ and $y\geq 1$. Then,
$d((1,1),(x,y))=|x-1|+|y-1|=x+y-2$. Therefore,
$a_{(1,1)}(x,y)=\frac{x+y-2}{s(1,1)}$ and
$a_{(x,y)}(1,1)=\frac{x+y-2}{s(x,y)}$. Now, observing that we can
calculate $s(1,1)$ as $$s(1,1)=\sum_{i=h+1}^{n-1} (i+1)\cdot i^{-r} +
\sum_{i=0}^{n-2} (n-1-i)\cdot (n+i)^{-r}$$ and using  expression
(\ref{eq:cw}) for $c_w$, the result follows.
\end{proof}

We are now ready to state our main result.
\begin{theorem}
For $h<n-1$ the capacity of a Kleinberg small-world network
graph lies, with high probability, in the interval
$\left[M,(1+\epsilon)c_w\right]$.
\label{th:kleinberg}
\end{theorem}

\begin{proof}
Using \lemmaref{lemma:mincut2} and \corolref{cor:karger}, we have
that, with high probability, $c_s \in
\left[(1-\epsilon)c_w,(1+\epsilon)c_w\right].$ A tighter lower
bound can be obtained for $c_s$ as follows. Each node has a number
of initial edges, determined by $h$, and $q$ additional shortcut
edges. The nodes with less initial edges are obviously the corner
nodes, which exhibit  $\frac{h(h+3)}{2}$ initial connections.
Therefore, we have that $c_s\geq \frac{h(h+3)}{2}+q,$ and the result
follows.
\end{proof}

\begin{figure}[h!]
\centering
\includegraphics[width=9.5cm]{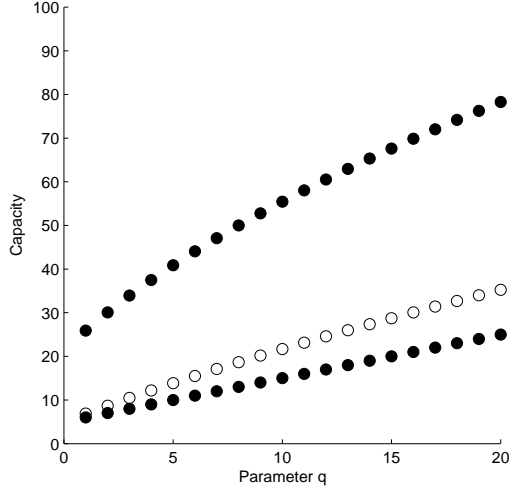}\vspace{-0.3cm}
\caption{Bounds for the capacity of Kleinberg small-world network
for $n=80$ (i.e.~1600 nodes), $h=2$, $r=2$ e $d=1$, and different
values of the shortcut parameter $q$. The white dots represent the
expected value of the capacity and the dark dots represent the
bounds computed according to \theorref{th:kleinberg}. }
\label{fig:kleinberg_sim}
\end{figure}

The bounds of \theorref{th:kleinberg} are illustrated in
\fig{fig:kleinberg_sim}.

\subsection{Capacity Bounds for Navigable Small-World Rings}

As we have seen, Kleinberg's model exhibits corner's effects in terms of
capacity. With the goal of overcoming this problem, we defined a new class of
small-world
networks, the navigable small-world ring (see \definref{def:navi} in
\secref{sec:swm}), whose navigability is proven in the appendix. Now we study
the capacity of this class of networks by proving the following result.

\begin{theorem}
\label{th:mymodel}
With high probability, the capacity of the navigable
small-world ring has a value in the interval
$\left[ \max \left\{k,(1-\epsilon)c_w) \right\},(1+\epsilon)c_w)\right]$, with
$\epsilon=\sqrt{2(d+2)\ln(n)/c_w}$ and $$c_w=k+2^{rq+1}s^{-q}
q(1+a_n)(n-a_n)^{-r} \left(2^{-r}s-(n-a_n)^{-r}\right)^{q-1}+
4qs^{-q}\cdot \sum_{i=\frac{k}{2}+1}^{\frac{n-a_n}{2}-1}
i^{-r}\left(s-i^{-r}\right)^{q-1}
$$
with $s=(1+a_n)\cdot \left( \frac{n-a_n}{2} \right)^{-r}+2\cdot
\sum\limits_{i=\frac{k}{2}+1}^{\frac{n-a_n}{2}-1} i^{-r}$,
where $a_n=\frac{1-(-1)^{n}}{2}$.
\end{theorem}

\begin{proof}
Consider the fully connected graph $G_w=(V_L, E)$ associated to the navigable
small-world graph. The task is to determine the weights of the edges of $G_w$.
The edges $e\in E_L$ have weight $1$, because we never remove them. Now, notice
that the ring distance between two nodes does not depend on which node is
numbered first. It is therefore correct to state that all the nodes have the
same number of nodes at a distance $d$. Therefore, we have that the normalizing
constants are equal, for all nodes: $s_i=s_j, \forall i,j$. Let $s=s_i$. We also
have that the weight of each edge only depends on the distance between the nodes
that it connects. Therefore, it is sufficient to determine the weights of the
edges of a single node, say node $1$.

First, we must compute the normalizing constant $s$. We must distinguish between
two different situations: even $n$ or odd $n$. If $n$ is even, it is not
difficult to see that there is a single node that maximizes the distance to node
$1$. That node is node $\frac{n}{2}+1$, and we have that $d\left(1,\frac{n}{2}+1
\right)=\frac{n}{2}$. For distances $d<\frac{n}{2}$, there are two nodes at
distance $d$ to node $1$. Therefore, if $n$ is even, we have that
$$s=\left( \frac{n}{2} \right)^{-r}+2\cdot
\sum_{i=\frac{k}{2}+1}^{\frac{n}{2}-1} i^{-r}.$$

When $n$ is odd, it is also easy to see that there are two nodes that maximize
the distance to node $1$: nodes $\frac{n+1}{2}$ and $\frac{n+3}{2}$, with the
maximum distance being $\frac{n-1}{2}$. Therefore, if $n$ is odd, we have that

$$s=2\cdot \sum_{i=\frac{k}{2}+1}^{\frac{n-1}{2}} i^{-r}.$$

Now, just notice that $a_n=\frac{1-(-1)^n}{2}$ is equal to $0$ if $n$ is even,
and it is equal to $1$ if $n$ is odd. Therefore,

$$s=(1+a_n)\cdot \left(
\frac{n-a_n}{2} \right)^{-r}+2\cdot \sum_{i=\frac{k}{2}+1}^{\frac{n-a_n}{2}-1}
i^{-r}.$$

Consider a node that is not initially connected to node $1$, say node $i$. The
edge $(1,i)$ can be added in two different trials: one for node $1$ and another
for node $i$. Because we do not consider multiple edges, these two trials are
mutually exclusive. Therefore, the weight of the edge $(1,i)$ is the sum of the
probabilities of adding this edge when looking at node $1$ and when looking at
node $i$. Because the normalizing constant is the same for all nodes, these two
probabilities are equal. This way, let us focus on node $1$. The trial ``add
edge $(1,i)$" follows a Binomial distribution, with $q$ Bernoulli experiences
and with success probability $p=\frac{d(1,i)^{-r}}{s}$. Therefore, the
probability of adding edge $(1,i)$ when considering node $1$ is
$qp\cdot(1-p)^{q-1}$. Therefore, the weight of the edge $(1,i)$ is
$$w(1,i)=2q\cdot \frac{d(1,i)^{-r}}{s}\cdot \left(
1-\frac{d(1,i)^{-r}}{s} \right)^{q-1}.$$

We have seen that all the nodes have the same number of nodes at a distance $d$.
We also have that all the edges in the ring lattice have unitary weight. Based
on these two observations and the fact that $G_w$ is a fully connected graph, it
is clear that the global minimum cut in $G_w$, denoted $c_w$, is a cut in which
one of the partitions consists of a single node, say node $1$. Thus,  we may
write
\begin{eqnarray*}
c_w &=& k+\sum_{i\in N_1} w(1,i)\\
&=& k+2(1+a_n)q\frac{\left(\frac{n-a_n}{2}\right)^{-r}}{s}\left(
1- \frac{\left(\frac{n-a_n}{2}\right)^{-r}}{s}
\right)^{q-1} +2\cdot \sum_{i=\frac{k}{2}+1}^{\frac{n-a_n}{2}-1}
2q\frac{i^{-r}}{s}\left(1-
\frac{i^{-r}}{s} \right)^{q-1}\\
&=& k+2^{rq+1}s^{-q} q(1+a_n)(n-a_n)^{-r}
\left(2^{-r}s-(n-a_n)^{-r}\right)^{q-1}+
4qs^{-q}\cdot \sum_{i=\frac{k}{2}+1}^{\frac{n-a_n}{2}-1}
i^{-r}\left(s-i^{-r}\right)^{q-1}
\end{eqnarray*}
Now, using \corolref{cor:karger} and noticing that, because we only add new
edges to the initial $k$-connected ring lattice and this lattice has capacity
$k$, the capacity can only be greater than $k$, we obtain the desired bounds.
\end{proof}

\begin{figure}[h!]
\centering
\includegraphics[width=9.5cm]{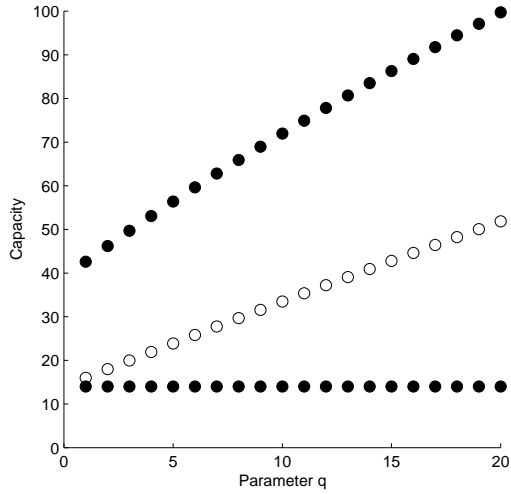}
\caption{Bounds for the capacity of a navigable small-world ring for
$n=1600$, $k=14$, $r=1$ e $d=1$, and different values of the shortcut parameter
$q$. The white dots represent the expected value of the capacity and the dark
dots represent the bounds computed according to \theorref{th:mymodel}.}
\label{fig:mymodelbounds}
\end{figure}

The result is  illustrated in \fig{fig:mymodelbounds}.

\section{Concluding Remarks}
\label{sec:conclusions}

We studied the max-flow min-cut capacity of four fundamental
classes of small-world networks. Using classical network flow
arguments and concentration results from random sampling in
graphs, we provided bounds for both standard and navigable
small-world networks with added shortcuts, and also for
Kleinberg's model. In addition, we presented a tight result for
small-world networks with rewiring, which permits the following
interpretation: {\it With high probability, rewiring does not
alter the capacity of the network}. This observation is not
obvious, because we can easily find ways to rewire the ring
lattice in order to obtain, for instance, a {\it bottleneck}. But,
according to the previous results, such instances occur with very
low probability.

In~\cite{kleinberg:00}, Kleinberg explains that, in order to
obtain a probability distribution, $d((u_1,u_2),(v_1,v_2))^{-r}$
should be divided by $\sum _{(v_1,v_2)\neq (u_1,u_2)}
[d((u_1,u_2),(v_1,v_2))]^{-r}.$ As we have shown, the previous
expression is not an  accurate normalizing constant for our work, because we
consider undirected edges. Then, the
candidates for new connections from node $(u_1,u_2)$ are {\it not
all} the nodes of the base lattice, but {\it only} those nodes
that are initially not connected to node $(u_1,u_2)$.

Possible directions for future work include tighter capacity
results, extensions to other classes of small-world networks (e.g.
weighted models and those used in peer-to-peer
networks~\cite{Manku:04}), and understanding if and how
small-world topologies can be exploited in the design of
capacity-attaining network codes and distributed network coding
algorithms. At a more conceptual level, we are intrigued by the
possibility that the notion of capacity may help us answer a very
central question: {\it why} small-world topologies are ubiquitous
in real-world networks.

\appendix
\section*{Proof of Navigability of the Small-World Ring}
\label{sec:app}

As we have discussed in \secref{sec:intro}, in the context of
communication networks, we would like that not only short paths
exist between any pair of nodes, but also that paths can easily be
found using only local information. Kleinberg,
in~\cite{kleinberg:00}, showed that this {\it navigability}
property is absent in the initial models of small-world networks,
from Watts and Strogatz. This way, Kleinberg felt the need to
introduce a new model, the model defined in \definref{def:klein},
that captures this fundamental property. In his work, Kleinberg
uses the idea of a {\it decentralized algorithm} to study the
navigability property.

\begin{definition}
Consider a graph $G$ with an underlying metric $\delta_G$. A {\it
decentralized algorithm} in $G$ is an algorithm with the goal of
sending a message from a source to a destination, with the
knowledge, at each step, of the underlying metric, the position of
the destination, and the contacts of the current message holder
and of all the nodes seen so far.
\end{definition}

\begin{definition}
A {\it greedy decentralized algorithm} is a decentralized
algorithm operating greedily: at each step, it sends the message
to the contact of the current message holder that is closer (in
the sense of the underlying metric) to destination.
\end{definition}

In~\cite{kleinberg:00}, Kleinberg proved that the models presented
by Watts and Strogatz do not admit efficient decentralized
algorithms, in constrast with his model:

\begin{theorem}[From~\cite{kleinberg:00}]
\label{th:kleinberg2} For $r=2$, there is a constant $\alpha_2$,
independent of $n$, so that the expected delivery time of a greedy
decentralized algorithm in a Kleinberg network is at most
$\alpha_2\cdot \log^2(n)$.
\end{theorem}

\begin{theorem}[From~\cite{kleinberg:00}]
\label{th:kleinberg3} $_{}$
\begin{enumerate}
\item Let $0\leq r <2$. There is a constant $\alpha_r$, depending on $p$, $q$,
$r$, but independent of $n$, so that the expected delivery time of any
decentralized algorithm in a Kleinberg network is at least $\alpha_r \cdot
n^{(2-r)/3}$.
\item Let $r>2$. There is a constant $\alpha_r$, depending on $p$, $q$, $r$, but
independent of $n$, so that the expected delivery time of any decentralized
algorithm in a Kleinberg network is at least $\alpha_r \cdot n^{(r-2)/(r-1)}$.
\end{enumerate}
\end{theorem}

\theorref{th:kleinberg2} shows that, in fact, a Kleinberg network
is navigable, while \theorref{th:kleinberg3} shows that the models
from Watts and Strogatz are not navigable, because this is the
case when we consider uniformly chosen shortcuts, therefore
corresponding to $r=0$.

The next theorem shows that a navigable small-world ring is,
indeed, navigable, in the sense that the expected delivery time of
a decentralized algorithm is logarithmic. The proof is essentially
based on the proof of \theorref{th:kleinberg2} presented by
Kleinberg.

\begin{theorem}
For $r=1$, the expected delivery time of a greedy decentralized
algorithm in a navigable small-world ring is at most
$\frac{\ln^2 (2n)}{\ln(2)}$.
\end{theorem}

\begin{proof}
First, we need to show that $\sum_{u\notin N_v} d(u,v)^{-1}$ is
uniformly bounded. For even $n$, it is not difficult to see that
there is a single node that maximizes the distance to node $1$.
That node is node $\frac{n}{2}+1$, and we have that
$d\left(1,\frac{n}{2}+1 \right)=\frac{n}{2}$. For distances
$d<\frac{n}{2}$, there are two nodes at distance $d$ to node $1$.
Therefore, if $n$ is even, we have that
$$\sum_{u\notin N_v} d(u,v)^{-1} = \left( \frac{n}{2}\right)^{-1} + 2\cdot
\sum_{i=\frac{k}{2}+1}^{\frac{n}{2}-1}i^{-1} \leq 2\cdot
\sum_{i=\frac{k}{2}+1}^{\frac{n}{2}}i^{-1}.$$

When $n$ is odd, it is also easy to see that there are two nodes
that maximize the distance to node $1$: nodes $\frac{n+1}{2}$ and
$\frac{n+3}{2}$, with the maximum distance being $\frac{n-1}{2}$.
Therefore, if $n$ is odd, we have that $$ \sum_{u\notin N_v} d(u,v)^{-1} =
2\cdot
\sum_{i=\frac{k}{2}+1}^{\frac{n-1}{2}} i^{-1}.$$

Therefore, we have that $\forall n\in$\NN,
\begin{eqnarray*}
\sum_{u\notin N_v} d(u,v)^{-1} &\leq& 2\cdot
\sum_{i=\frac{k}{2}+1}^{\left\lfloor \frac{n-1}{2}\right\rfloor} i^{-1} \\
&\leq& 2\cdot \sum_{i=1}^{\left\lfloor \frac{n-1}{2}\right\rfloor} i^{-1}\\
&\leq& 2+2\ln \left( \frac{n}{2} \right)\\
&\leq& 2\ln (2n).
\end{eqnarray*}

For $j>0$, we say that the decentralized algorithm is in phase $j$
if the distance between the current message holder and the
destination is $d$ such that $2^j<d\leq 2^{j+1}$. We say that the
algorithm is in phase $0$ if the distance between the current
message holder and the destination is at most $2$. Because the
maximum distance in the ring-lattice is at most $\frac{n}{2}$, we
have that $j\leq \log \left( \frac{n}{2}\right)$.

Now, suppose that we are in phase $j$ and the current message
holder is node $u$. The task is to determine the probability of
phase $j$ ending in this step. Let $B_j$ be the set of nodes
within lattice distance $2^j$ of the destination. Phase $j$ ending
in this step means that $u$ chooses a long-range contact $v\in
B_j$. Each node $v\in B_j$ as probability of being chosen as
long-range contact of $u$ at least
$$ \frac{\left( 2^j\right)^{-1}}{\sum_{v\notin N_u} d(u,v)^{-1}}\geq
\frac{1}{2^{j+1}\cdot \ln (2n)}.$$
We have that the number of nodes in $B_j$, denoted by $|B_j|$,
verifies
$$|B_j|=1+2\cdot \sum_{i
=1}^{2^j} i \geq 2^{2j}.$$
Therefore, with $A$ denoting the event
``Phase $j$ ends in this step'', we have that
$$\prob(A) \geq \frac{2^{2j}}{2^{j+1}\cdot \ln (2n)} = \frac{2^{j-1}}{\ln (2n)}
\geq \frac{1}{\ln (2n)}.$$

Let $N_j$ be the number of steps spent in phase $j$. Now, we must
compute the expected value of $N_j$. Notice that the maximum
number of steps spent in phase $j$ is the number of nodes at
distance of the destination $d$ such that $2^j<d\leq 2^{j+1}$,
which is

\begin{eqnarray*}
m &=& 2\cdot \sum_{i=2^j}^{2^{j+1}-1} i \\&=& 2\cdot \left(
\sum_{i=1}^{2^{j+1}-1} i - \sum_{i=1}^{2^j-1} i \right)\\&=&
2^{j+1}(2^{j+1}-1)-2^j(2^j-1)\\&\leq& 2^{2j+2} .
\end{eqnarray*}

Therefore, the expected value of $N_j$ verifies the following:

\begin{eqnarray*}
E\left( N_j \right)&=& \sum_{i=1}^{m} \prob (N_j \geq i)\\
&\leq& \sum_{i=1}^{2^{2j+2}} \prob (N_j \geq i)\\
&\leq& \sum_{i=1}^{2^{2j+2}} \left(  1- \frac{1}{\ln (2n)}\right)^{i-1}\\
&\leq& \sum_{i=1}^{\infty} \left(  1- \frac{1}{\ln (2n)}\right)^{i-1}\\
&=& \ln(2n).
\end{eqnarray*}

Now, denoting by $N$ the total number of steps spent by the
algorithm, we have that
$$N=\sum_{i=0}^{\log \left( \frac{n}{2}\right)} N_j.$$
Therefore, by linearity of the expected value, we have that

\begin{eqnarray*}
E\left( N \right)&=& \sum_{i=0}^{\log \left( \frac{n}{2}\right)} E\left(
N_j\right)\\
&\leq& \left(1+\log \left( \frac{n}{2}\right)\right)\cdot \ln(2n)\\
&=&\left(\log(2)+\log \left( \frac{n}{2}\right)\right)\cdot \ln(2n)\\
&=&\log(n)\cdot \ln (2n)\\
&=&\frac{\ln(n)\cdot \ln(2n)}{\ln(2)}\\
&\leq& \frac{\ln^2 (2n)}{\ln(2)}
\end{eqnarray*}

\end{proof}

\bibliographystyle{IEEE}
\bibliography{bb}

\begin{thebibliography}{10}

\bibitem{Costa-Barros:06a}
Rui~A. Costa and J.~Barros,
\newblock ``On the capacity of small world networks,''
\newblock in {\em Proceedings of the IEEE Information Theory Workshop}, March
  2006.

\bibitem{Costa-Barros:06b}
Rui~A. Costa and J.~Barros,
\newblock ``Network information flow in {\it navigable} small-world networks,''
\newblock in {\em Proceedings of the IEEE Workshop in Network Coding, Theory
  and Applications}, April 2006.

\bibitem{milgram:67}
S.~Milgram,
\newblock ``Phychology today,''
\newblock {\em Physical Review Letters}, vol. 2, pp. 60--67, 1967.

\bibitem{kleinfeld:02}
J.~S. Kleinfeld,
\newblock ``Could it be a big world after all?,'' Society, 2002.

\bibitem{yamamoto:92}
T.B. Achacoso and W.S. Yamamoto,
\newblock {\em AY's Neuroanatomy of C. elegans for Computation},
\newblock CRC Press, 1992.

\bibitem{martinez:00}
R.J. Williams and N.D. Martinez,
\newblock ``Simple rules yield complex food webs,''
\newblock {\em Nature}, vol. 404, pp. 180--183, 2000.

\bibitem{westbrook:98}
James Abello, Adam~L. Buchsbaum, and Jeffery Westbrook,
\newblock ``A functional approach to external graph algorithms,''
\newblock in {\em ESA '98: Proceedings of the 6th Annual European Symposium on
  Algorithms}, London, UK, 1998, pp. 332--343, Springer-Verlag.

\bibitem{newman:01}
M.E.J. Newman,
\newblock ``The structure of scientific collaboration networks,''
\newblock in {\em Proc. Natl Acad. Sci.}, 2001, vol.~98, pp. 404--409.

\bibitem{broder:00}
A.~Broder,
\newblock ``Graph structure in the web,''
\newblock {\em Comput. Netw.}, vol. 33, pp. 309--320, 2000.

\bibitem{watts:98}
Duncan~J. Watts and Steven~H. Strogatz,
\newblock ``Collective dynamics of 'small-world' networks,''
\newblock {\em Nature}, vol. 393, no. 6684, June 1998.

\bibitem{NewmanW:99}
Mark E.~J. Newman and Duncan~J. Watts,
\newblock ``Scaling and percolation in the small-world network model,''
\newblock {\em Physical Review E}, vol. 60, no. 6, pp. 7332--7342, December
  1999.

\bibitem{strogatz:01}
Steven~H. Strogatz,
\newblock ``Exploring complex networks,''
\newblock {\em Nature}, vol. 410, pp. 268--276, March 8 2001.

\bibitem{Lehman:04}
April~Rasala Lehman and Eric Lehman,
\newblock ``Complexity classification of network information flow problems,''
\newblock in {\em Proceedings of the 15th annual ACM-SIAM symposium on Discrete
  algorithms}, Philadelphia, PA, USA, 2004, pp. 142--150, Society for
  Industrial and Applied Mathematics.

\bibitem{ford:62}
L.R. Ford and D.R. Fulkerson,
\newblock {\em Flows in Networks},
\newblock Princeton University Press, Princeton, NJ, 1962.

\bibitem{BarrosS:06}
J.~Barros and S.~D. Servetto,
\newblock ``Network information flow with correlated sources,''
\newblock {\em IEEE Transactions on Information Theory}, vol. 52, no. 1, pp.
  155--170, January 2006.

\bibitem{AhlswedeCLY:00}
R.~Ahlswede, N.~Cai, S.-Y.~R. Li, and R.~W. Yeung,
\newblock ``{Network Information Flow},''
\newblock {\em IEEE Transactions on Information Theory}, vol. 46, no. 4, pp.
  1204--1216, 2000.

\bibitem{Borade:02}
S.~Borade,
\newblock ``{Network Information Flow: Limits and Achievability},''
\newblock in {\em Proc. IEEE Int. Symp. Inform. Theory (ISIT)}, Lausanne,
  Switzerland, 2002.

\bibitem{LiYC:03}
S.-Y.~R. Li, R.~W. Yeung, and N.~Cai,
\newblock ``{Linear Network Coding},''
\newblock {\em IEEE Trans. Inform. Theory}, vol. 49, no. 2, pp. 371--381, 2003.

\bibitem{KoetterM:03}
R.~Koetter and M.~M\'edard,
\newblock ``{An Algebraic Approach to Network Coding},''
\newblock {\em IEEE/ACM Trans. Networking}, vol. 11, no. 5, pp. 782--795, 2003.

\bibitem{ramamoorthy:03}
Aditya Ramamoorthy, Jun Shi, and Richard~D. Wesel,
\newblock ``On the capacity of network coding for random networks,''
\newblock in {\em Proc. of 41st Allerton Conference on Communication, Control
  and Computing}, Allerton, Illinois, October 2003.

\bibitem{GuptaK:00}
P.~Gupta and P.~R. Kumar,
\newblock ``The capacity of wireless networks,''
\newblock {\em IEEE Trans. Inform. Theory}, vol. 46, no. 2, pp. 388--404, March
  2000.

\bibitem{PerakiS:05}
C.~Peraki and S.~D. Servetto,
\newblock ``{On Multiflows in Random Unit-Disk Graphs, and the Capacity of Some
  Wireless Networks},'' Submitted to the IEEE Trans. Inform. Theory, March
  2005. Available from {\tt http://cn.ece.cornell.edu/}.

\bibitem{mendes:03}
S.N. Dorogovtsev and J.F.F. Mendes,
\newblock {\em Evolution of Networks: From Biological Nets to the Internet and
  WWW},
\newblock Oxford University Press, 2003.

\bibitem{helmy:03}
A.~Helmy,
\newblock ``Small worlds in wireless networks,''
\newblock {\em IEEE Communications Letters}, vol. 7, no. 10, pp. 490--492,
  October 2003.

\bibitem{reznik:03}
Alex Reznik, Sanjeev~R. Kulkarni, and Sergio Verd\'u,
\newblock ``A small world approach to heterogeneous networks,''
\newblock {\em Communication in Information and Systems}, vol. 3, no. 4, pp.
  325--348, 2004.

\bibitem{Dixit:05}
S.~Dixit, E.~Yanmaz, and O.~K. Tonguz,
\newblock ``{On the design of self-organized cellular wireless networks},''
\newblock {\em IEEE Communications Magazine}, vol. 43, no. 7, pp. 86--93, 2005.

\bibitem{Manku:04}
G.~S. Manku, M.~Naor, and U.~Wieder,
\newblock ``Know thy neighbor's neighbor: The power of lookahead in randomized
  p2p networks,''
\newblock in {\em Proceedings of the 36th ACM Symposium on Theory of
  Computing}, 2004.

\bibitem{kleinberg:00}
Jon Kleinberg,
\newblock ``The small-world phenomenon: an algorithm perspective,''
\newblock in {\em Proceedings of the 32th annual ACM symposium on Theory of
  Computing}, New York, NY, USA, 2000.

\bibitem{karger:94}
David~R. Karger,
\newblock ``Random sampling in cut, flow, and network design problems,''
\newblock in {\em STOC '94: Proceedings of the twenty-sixth annual ACM
  symposium on Theory of computing}, New York, NY, USA, 1994.

\end{thebibliography}

\end{document}